# X-ray tomographic intervention guidance: Towards real-time 4D imaging

Technical development

**Sönke Bartling**
**Marc Kachelrieß**

PD Dr. med. Sönke Bartling (corresponding author)
soenkebartling@mailbox.org
+49 151 585 85 585

Prof. Dr. rer. nat. Marc Kachelrieß
m.kachelriess@dkfz.de

**X-ray and Computed Tomography**
Dept. of Medical Physics in Radiology

**German Cancer Research Center**
Im Neuenheimer Feld 280
69120 Heidelberg
Germany

# Abstract

Implementation of real-time, continuous, and three-dimensional imaging (4D intervention guidance) would be a quantum leap for minimally-invasive medicine. It allows guidance during interventions by assessing the spatial position of instruments continuously in respect to their surroundings. Recent research showed that it is feasible using X-ray and novel tomographic reconstruction approaches. Radiation dose stays within reasonable limits. This article provides abstractions and background information together with an outlook on these prospects. There are explanations of how situational awareness during interventions is generated today and how they will be in future. The differences between fluoroscopically and CT-guided interventions are eluted to within the context of these developments. The exploration of 'uncharted terrain' between these current methods is worth pursuing. Necessary image quality of 4D intervention guidance varies relevantly from that of standard computed tomography. Means to analyze the risk-benefit ratio of 4D intervention guidance are provided and arguments for gantry-based setups vs C-arm based setups are given. Due to the lack of moving organs, neuro-interventions might be the first field in which 4D intervention guidance might become available, however, registration and fusion algorithms might make it applicable in complex whole-body interventions such as aortic valve replacement soon thereafter.

# Introduction

**Scope of This Review**

Current developments in technology may allow for far-reaching changes in minimally-invasive medicine. For the first time in the long-standing tradition of X-ray guided interventions, it seems possible that continuous, 4D real-time imaging during minimally-invasive procedures may be feasible [1–3] ('4D' in this context refers to three spatial dimensions and one time dimension). In this review, we will elucidate upon the underlying basis of these developments, explain concepts, build abstractions and will hopefully provide insightful points of view, and discuss the potential implications for the future of intervention guidance. By envisioning developments to come, potential obstacles and challenges can be identified, which initiate discussions that might guide the industry and the interventional radiology communities in the right directions and, possibly avoid potential dead ends in developments. Furthermore, long standing views on how to technically setup intervention guidance environments are challenged and discussed.

This review focuses on minimally-invasive interventions which make use of catheters and (biopsy) needles. Interventions in a much broader sense such as orthopedic or trauma surgeries, which may also be based on X-ray (through both fluoroscopy as well as CT guidance), are not considered in this essay.

**Introduction to Intervention Guidance**

Image guidance of interventions inside of the human body is a key concept in minimally-invasive medicine. While the direct line of sight is inhibited, interventions are guided by means of imaging which provides the position of intervention instruments in respect to the patient´s anatomy. Additionally, imaging methods are used to assess treatment effects during procedures.

Many different imaging methods exist, all with inherently different advantages and disadvantages. Of these, X-ray guidance is still the prevailing method. It has a long history of use because X-rays provide a good trade-off between costs, harm, and imaging characteristics. X-rays are of sufficient penetration depth, their contrast effects are largely independent from the surrounding structures and they provide an undistorted geometry. Despite remarkable developments in the field of MRI guided interventions, X-ray guidance has not lost significant ground. This is most likely due to the installation requisites for providing image guidance with MRI which are high and, despite decades of intensive research, cannot, as of yet, be justified by the advantages of MRI guidance [4,5]. However, considerable efforts in the research for MRI based intervention guidance, stem from unmet needs in current intervention guidance [6]. Moreover, they are symptomatic of an urge to further develop minimally-invasive medicine towards an ideal image guidance method that will be outlined in the following section.

**Theoretical Consideration: Ideal Intervention Guidance Method and the Four Sources of Situational Awareness During Interventions**

Ideally, a method for intervention guidance would provide a continuous and spatial depiction of the patient's anatomy and pathology in addition to displaying the position of the intervention instrument. All three of these benefits create the necessary *situational awareness*. More so, an ideal method would perform this in real-time, meaning that there is no significant delay between changing the intervention instrument's position and the display of the current spatial situation. Obviously, this should be performed in a manner that does no harm to the patient, is inexpensive, and does not hinder access to the patient. 3D information generated should be highly accessible and displayed meaningfully with minimal user-interactions.

Evidently, the closer that one gets to this ideal situation of intervention guidance, the higher will be the gains generated by it. Interventions will be more accurate, complications will be reduced, and the risk to benefit ratio will be optimized. Training requirements for interventionalists would be minimized, and more and more complex procedures could be invented and developed. This would go hand in hand with the development of novel intervention instruments and implants.

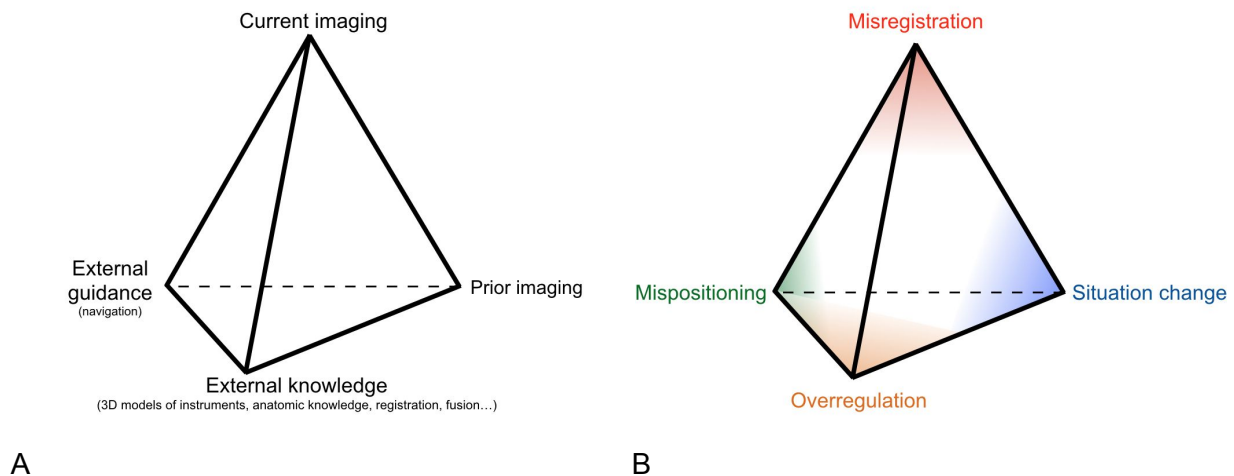

**Figure 1. Situational awareness during intervention guidance** is created by four (non-human) factors (A) that are prone to specific errors and misconceptions (B). Within this article, we will focus upon novel combinations of tomographic *current imaging*, *prior imaging* and *external knowledge* which may ultimately make 4D intervention guidance possible.

Efforts to improve current intervention guidance are large. These efforts consists of integrating additional sources of information in order to increase the situational awareness.

The combination of navigation guidance with interventional imaging [7,8] is one way that this is achieved. Intraprocedural navigation has so far found wide application within cardiologic interventions [9]. In other fields, the benefits do not seem to have a worthy trade-off for the additional preparations and expenditures, since navigation is not widely employed.

Co-registration and fusion of 3D overlays, generated by prior imaging, to the projective imaging [10][11] serve as another means to increase situational awareness.

Both methods provide additional information, but still rely largely upon standard projective imaging to update the surrounding anatomy and the relationship of intervention instruments. Therefore, they are prone to misregistration and misalignment errors [12]. To minimize these errors, 3D images can be acquired repetitively throughout the intervention.

Furthermore, prior knowledge about instruments and anatomy were introduced into guidance by numerous approaches in order to compensate for shortcomings in imaging methods [13]. However, these methods might be prone to over-regulation and might provide a false margin of safety for the interventionalist to work within.

To sum up, besides human factors such as experience, knowledge, sensory/haptic feedback, etc., the situational awareness of the interventionalist is generated by four different themes (Fig. 1 a):

- Prior imaging (e.g. pre-interventional CT or MRI, intraprocedural cone-beam CT)
- external knowledge (e.g. anatomy, atlases, models)
- external guidance (e.g. navigation system)
- current imaging (e.g. fluoroscopy, recent tomographic imaging)

Co-integration comes with certain aforementioned limitations (Fig. 1 b). The herein envisioned 4D intervention guidance is no exception. Rather, it represents a novel integration of prior imaging, current imaging, and a special form of external knowledge. As with all novel methods, this one comes with new opportunities, but also new caveats, as will be elucidated on in this paper.

**Current Status Quo of X-ray Intervention Guidance and Ongoing Developments**

To understand the potential of 4D intervention guidance, one should take a closer look at the ways in which X-ray interventions are guided today. Currently, interventions are done in two different kinds of environments: the CT suite and Cath Lab. For a comparison, see (Tab. 1).

**Table 1. Comparison between guidance in the CT suite and Cath Lab.**

|  | CT suite | Cath Lab |
| --- | --- | --- |
| Setup | Gantry based | C-arm |
| Imaging setup | Diagnostic multislice-CT | Flat-panel based fluoroscopy |
| Provided information | 3D data sets | 2D projective images |

| Timeliness | Still data sets at given time points in between manipulations | Continuous imaging |
| --- | --- | --- |
| Workflow | Movement guided by haptic feedback, then acquisition of 3D dataset, evaluation of situation, replanning manipulation, ab initio | Continuous imaging during manipulation, trial-and-error approaches in situations were spatial relationship is not clear |
| Procedures | Biopsies, punctures, drainages (PTCD), tumor ablations, … | Vascular interventions, drainages (incl. PTCD), TIPS, ... |

Both environments provide different kinds of guidance information that determine how procedures can be conducted. These are explained in the following points:
- Within the *CT suite*, interventions are performed using a step-and-shoot approach. This entails, the instruments are manipulated blindly, solely with haptic feedback, in between subsequent acquisitions of CT data sets. CT data sets provide the positions of the instruments and corrections can be planned. Continuous CT-imaging, known as "CT-fluoroscopy" was described [14], but it was never considered as a routine application method due to its high levels of radiation [15]. Interventions in the CT suite are usually done with a stiff needle, which is advanced into a specific area, after which catheters, biopsy instruments, or ablation instruments are introduced.
- Imaging within the *angiography suite* allows continuous and real-time imaging. Catheter based, intravascular interventions are the most prevalent kind of procedures, and a vast variety of interventions exist. Additionally, extravascular interventions are conducted. Needles are used to drain abscesses, as well as clogged gall ducts. Since the third dimension is missing in projective imaging, the needle is introduced using additional heuristics such as anatomic landmarks. By constantly injecting contrast media, the needle is then advanced or retracted along a path until contrast media is detected within the targeted structure (e.g. in gall ducts during PTCD).

4D intervention guidance would combine the best features of both worlds, 3D imaging that would be continuous at the same time.

**Increasing Availability of Cone-Beam CT Within the Cath Lab**
Currently, both previously described guidance methods are merging with respect to the setup. In the past years, all major vendors introduced C-arm systems that can rotate around the patient, acquiring projective images from a multitude of angles as they do so. From these projections, a standard cone-beam CT is reconstructed leading to a volumetric dataset, just as in a multi-slice CT scanner.

The image quality is different from that of standard multislice-CT, mostly owing to the different detector technology. The spatial resolution is usually higher, while the noise level is also increased and the soft-tissue contrast is lower than in multislice-CT. The consistency of the CT-numbers throughout the dataset is less homogeneous, and artifact behavior is different. In terms of radiation distribution and radiation protection, some significant differences exist when compared to standard CT, mostly resulting from the altered environment in which CT scanning is being used, and by the lack of a 360° rotation [16,17].

However, the image quality is completely sufficient not only for high contrast structures such as bone / background or contrast media filled vessels / background, but also for clinically relevant low contrast situations, such as blood / brain, abscesses / tissue and dilated bile ducts/liver tissue differentiation [18].

Cone-beam CT scanning within the Cath Lab was initially introduced as a replacement for diagnostic CT scans. Before its introduction, when a complication was suspected or the results of the intervention had to be evaluated, the patient had to be transported to the diagnostic CT suite, which caused significant logistic effort. If the intervention had to be continued, the patient then had to be moved back to the intervention room. With a cone-beam CT available within the Cath Lab, intracranial bleedings can be diagnosed, a therapy response can be evaluated, and the positions of drainages, coils, and bleedings can be confirmed, all with minimal logistical effort. From this cone-beam CT´s use cases / indications are currently evolving which will be elucidated upon in the following.

**Cone-beam-CT Inside the Cath Lab to Actually Guide Interventions**

The cointegration of cone-beam-CT with projective fluoroscopy allows for a combination of both aforementioned guidance methods: CT imaging provides a 3D overview, while fluoroscopic imaging provides a continuous imaging in between CT-scan to e.g. guide a needle (an example workflow is provided (Fig. 2)).

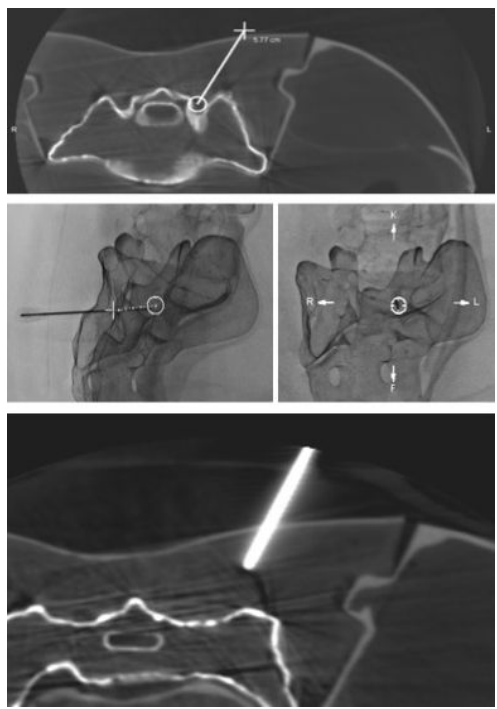

**Figure 2. Cointegration of fluoroscopic and CT guidance for intervention.** The trajectory is being planned with a 3D data set from a cone-beam-CT scan which was performed using the C-arm system (top), while the advancement of the needle is being monitored with fluoroscopy along a projection of the proposed pathway (middle). The C- arm can be alternated in between two purposefully preselected projections that can provide a good position to maintain control of the needle. If in doubt, a novel CT scan can be acquired to verify instrument positions (bottom).

For catheter interventions, cone-beam-CT is currently used with increasing frequency to guide the intervention. For example, CT-scans with contrast media that depict the vasculature with high contrast, are being overlaid upon the actual fluoroscopic guidance images [10,11]. Besides providing a `road map`, the 3D volume rendering of the vessel tree can be used to find projection positions that will maximize the guidance information by reducing the overlap of structures, which are currently of interest [19]. The C-arm is then automatically brought into the proposed position. Another use case of cone-beam-CT is to plan the position of implants in complex procedures [20,21]. Furthermore, it is used to verify the position of intervention instruments and implants. A remarkable example is the coiling of brain artery aneurysms, especially in stent protected aneurysm coiling. Here, it is very important to know the position of coils with respect to stent struts [22–24]. Many more cases for future use can be envisaged. One that seems to be close to clinical applicability is the planning of the reentry direction after crossing a stenosis subintimal.

The advantages of CT for actual guidance is easy to explain. CT provides an overlapping free depiction of the position and relationship of implants, and intervention instruments, as well the patient's anatomy and, moreover, many pathologies. Despite the obvious advantages, cone-beam CT is still relatively seldomly used during today's catheter interventions. The most likely reasons for this include:

*1. Customization to a guidance with CT*

Given the fact that catheter intervention guidance worked for decades without CT imaging, it is certainly a process of adaptation which is currently taking place. Thus, it may be assumed that CT imaging will be used more frequently in the future after a learning/adaptation curve.

*2. Radiation dose concerns*

CT scanning is generally perceived as a high-dose imaging modality. However, as always the case in therapeutic decision-making, the benefits must be weighed against potential harm. In the future it might become increasingly clear that the potential harm of flat-panel CT imaging during interventions could be completely outweighed by its benefits.

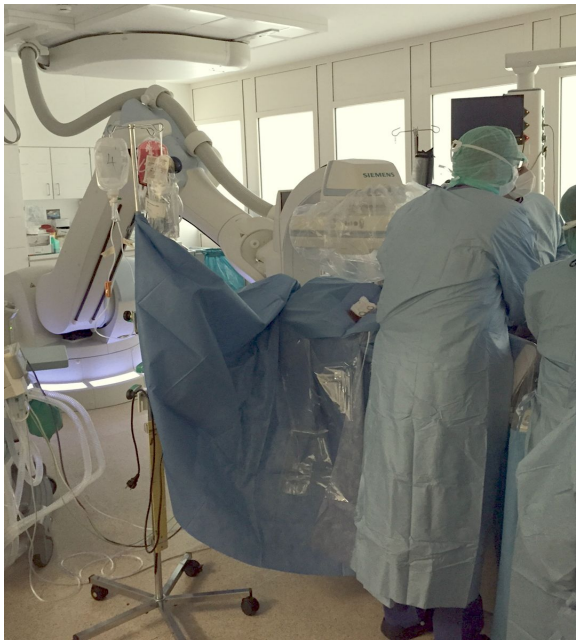

**Figure 3: Realistic situation within the Cath Lab during an intervention:** The rotational space for the C-arm is covered with wires, lines, etc., increasing the preparatory threshold to acquire a CT scan and therefore limiting the usability of CT scans for guidance.

*3. The considerable logistic efforts required to actually perform a CT scan*

A relevant effort is necessary to bring the imaging chain in position to acquire a tomographic data set and to clear the scanning trajectory. Usually, a tangle of lines, cables, infusions, and tubing blocks the trajectory pathway for the C-arm. Sterile fabrics as well as plastic covers limit the space and are in constant danger of becoming contaminated (Fig. 3). To allow a tomographic scan, many things must be rearranged carefully. Usually, this takes a considerable amount of time. A good management of space and anticipatory organization can reduce the effort required. In any case, interventions have to stop for a certain period of time in order to prepare and acquire the CT scan, during which no guidance information can be provided. This is especially perturbing in critical parts of the intervention - sometimes in those situations where a 3D image would be most useful.

## Current Development: Towards 4D Intervention Guidance

To sum up, the status quo of X-ray imaging in intervention guidance consists of a combination of fluoroscopy, biplane fluoroscopy, and computed tomography with an increasing tendency towards the application of tomographic imaging. Tomographic imaging complements the information provided by fluoroscopy at certain time points during the intervention.

**A Theoretical Consideration ...**
An interesting observation may be made if these guidance methods are seen as endpoints of a continuous spectrum of possible guidance methods. Fluoroscopy (incl. bi-plane fluoroscopy) could be considered to be one end and can be characterized by relatively low dosage. With this low dosage comes a relatively low amount of information content, almost entirely lacking spatial information in the third dimension. Computed tomography, on the other end, comes with a high information content, but needs time to be acquired. Contained in a semiquantitative/qualitative diagram which plots the information content (including spatial information) and radiation dose over the 'timeliness' of a method - as done in Figure 4 - results in a spectrum (blue) which ends are taken by the current methods (blue ends). The whole area in between those both extremes is 'unchartered terrain' (red). By exploring this uncharted terrain, one might discover novel and useful methods with interesting trade-offs between level of (spatial) information, sufficient timeliness, and dosage rates. The technologies discussed in the next paragraphs make this exploration possible by providing continuity between both extremes.

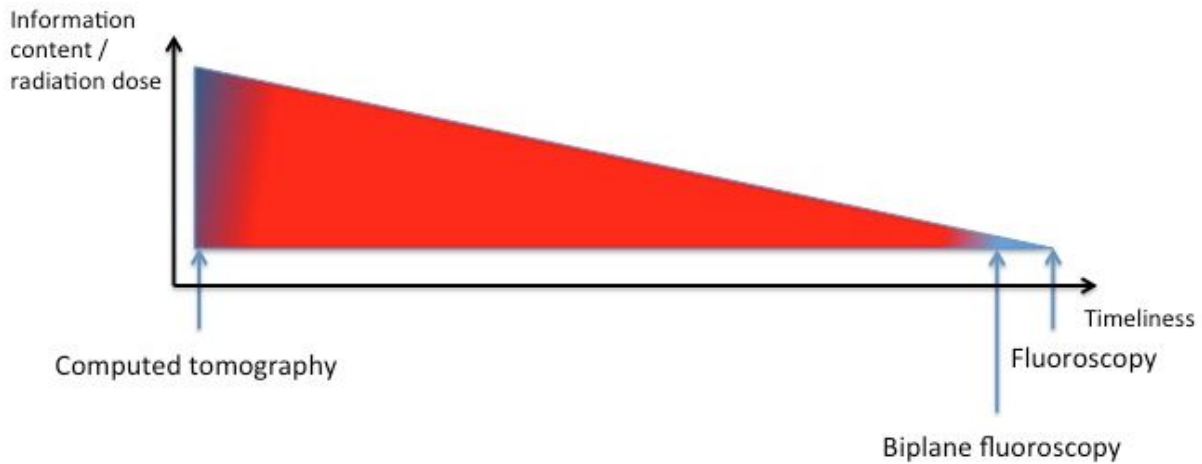

**Figure 4. Currently two methods are being used in X-ray intervention guidance, constituting the endpoints of a spectrum.** The red area is largely 'uncharted terrain' - with potentially useful, tomographic methods with new and interesting trade-offs between radiation dose and information content, which can result in innovative guidance methods [1]. The herein discussed methods allow the exploration of this terrain.

**Special Constraints of Tomographic Imaging in Intervention Guidance**
The requirements for tomographic imaging in intervention guidance are different from tomographic imaging in diagnostic imaging. These constraints can be interpreted as a kind of *external knowledge* as introduced in the introduction:

1. Same or very similar objects are being repeatedly scanned and the changes over time are small
2. Mostly high contrast structures are of interest (wires, catheters, contrast media filled vessels)
3. Absolute or relative consistency of the CT values is not necessary
4. Certain types of artefacts can be tolerated without losing the ability to provide sufficient guidance information
5. Typical prior knowledge (anatomy, structure of the instruments, etc.) is available which is specific to intervention guidance

All of these characteristics can be adapted to tomographic scanning and reconstruction in the special situation of intervention guidance. Resulting in the decrease of the necessary radiation dose, which may finally result in a continuous, tolerable 4D imaging, as will be elucidated upon in the following section.

**Novel Reconstruction Algorithms for 4D Intervention Guidance**

Standard tomographic reconstruction algorithms treat every volumetric dataset as an independent volume. Therefore, they cannot make use of the first of the abovementioned constraints, i.e. that changes over time are usually small.

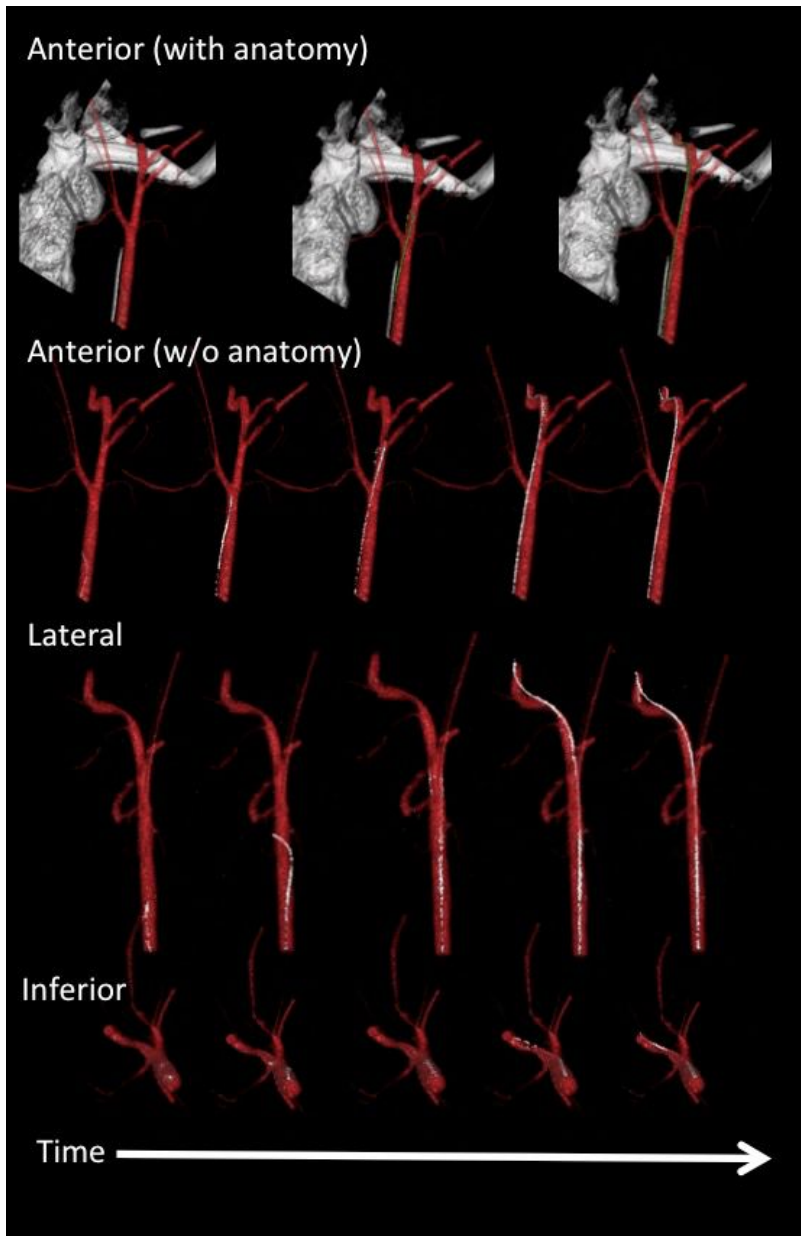

**Figure 5: Exemplary display of 4D intervention guidance from a living pig experiment.** The guidewire (white) can be assessed within the carotid artery (red) from all potential directions whereof three (anterior, lateral and inferior) are depicted herein. Exemplarily, surrounding anatomy (vertically oriented pig skull base) is provided in the anterior direction.

Furthermore, if certain criteria are unmet in standard tomographic reconstruction, e.g. the amount of available projection or radiation dose, certain artefacts occur that may make intervention guidance impossible.

Recently, many novel reconstruction schemes were employed which, when adequately used, provide interesting features by which the aforementioned constraints may be adequately explored. Many of the reconstruction approaches have been known to the reconstruction community long beforehand [25]. Namely, these are iterative reconstruction algorithms [26,27], algorithms that employ prior knowledge/models [28,29], and compressed sensing algorithms [30–33], to name a few examples. Many have become commonplace now in diagnostic imaging and thereby represent the novel state of the art. However, in interventional CT, only standard reconstruction algorithms have been employed as of yet.

By finding the right constraints and by identifying useful prior knowledge to be incorporated into the reconstruction algorithm, the amount of necessary projection could be reduced. This has been done, to an extent, so that real-time 4D imaging might become possible with a tolerable radiation dose essentially allowing 4D intervention guidance which has been shown recently (Fig. 5) [1–3].

In these algorithms, a *prior imaging* was combined with a low dose *current imaging* and the differences were reconstructed using an adapted compressed sensing scheme [1]. Two variations of this concept have been described in the literature [2,3]. Many more are thinkable which may result in interesting trade-offs (Fig 6.).

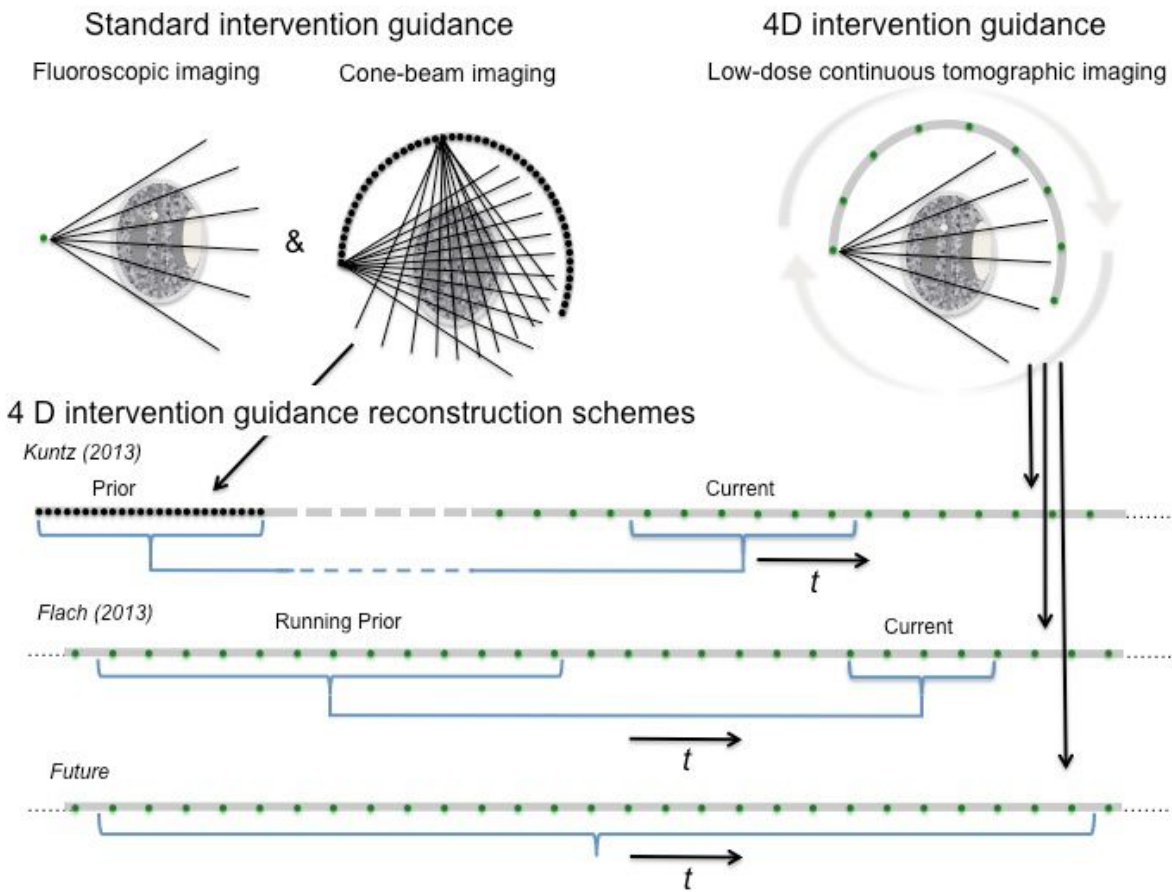

**Figure 6: Data acquisition and reconstruction schemes in standard and 4D intervention guidance.** Projection positions are symbolized by dots. Blue lines and brackets symbolize the data that is used for time frame reconstruction. Kuntz et al. combined data from a high-dose Prior scan (which is essentially same as a standard cone-beam CT scan) with continuously acquired, low-dose update (current) tomographic data. Flach et al. used low-dose data that was acquired anyway for a 'running Prior' to save dose and to compensate for bulk patient movements. It is expected that in the future reconstruction algorithms will make use of larger chunks of low-dose data to reconstruct most meaningful intervention guidance time frames.

# Discussion and outlook

**The Right Demand for Image Quality in 4D Intervention Guidance**

When thinking about the appropriate image quality of tomographic images for 4D intervention guidance, one must take the purpose of the imaging into account. Image quality is a function of the radiation dose.

Sufficient image quality for intervention guidance is much different from that of a diagnostic CT. Requisites for diagnostic images are defined in terms of objective features (noise level, signal to noise level, consistency of Hounsfield units, etc.) and subjective features (texture, sharpness, etc.), but these are entirely (not yet) defined for tomographic intervention guidance.

Tomographic reconstruction useful for intervention guidance needs to display the intervention instruments and the surrounding anatomy with the level of detail necessary for the actual situation within the intervention. A sufficient localization of the intervention instruments, with respect to the surrounding anatomy and other instruments might be very well deduced from tomographic images, although the quality of which is far below of that of diagnostic images. Noise, artefacts, and many more parameters might be much worse than in standard diagnostic CT images (Fig. 7). Furthermore, the image quality requirements may vary throughout the intervention, and may also depend upon the experience of the interventionalist.

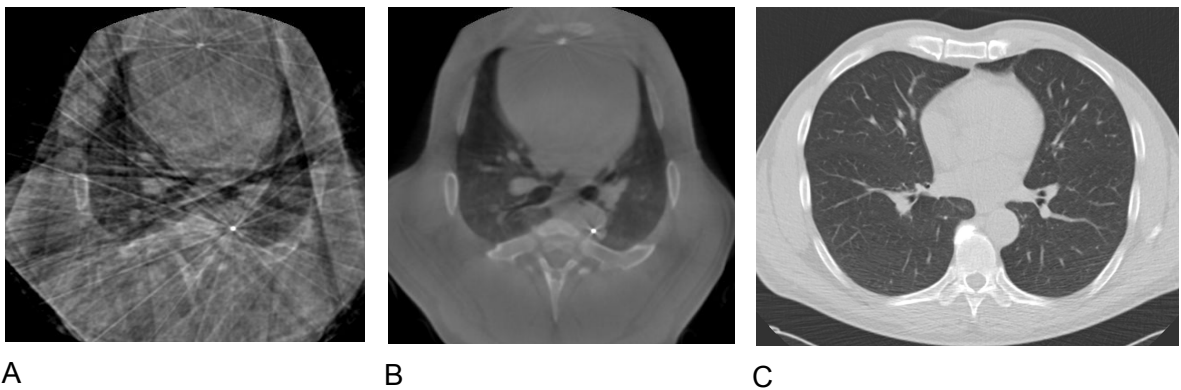

A  B  C

**Figure 7: Tomographic reconstruction for intervention guidance.** Image quality like the one in (A) is entirely sufficient for intervention guidance; this stands in stark contrast to conventionally scanned cone-beam CT data, which is depicted in (B) and standard, diagnostic multislice CT data (C). Despite streak artefacts, the intervention instruments (e.g. guidewires (white dot in the lower third (A,B)) are clearly depicted (A and B are examples from rabbits, C is from a human scan).

**Motion Compensation and Registration**

When different time points are combined into image reconstruction, as was done in the mentioned 4D intervention guidance schemes [1,2], movements within the datasets need to be considered [3].

While the proposed algorithms are inherently designed to incorporate the motion of interventional instruments into the 3D datasets, compensation for larger movements by the patient themself is not inherently fixed. This holds true for occasional motions of (a conscious) patient as well as physiological motions (breathing, heartbeat) which are constantly present during the intervention.

Bulk movements in 4D intervention guidance can be addressed using registration methods [3,34]. The head with all its rigid structures is an especially well suited field for this kind of motion compensation. In contrast, the thorax or the abdomen are much more complex; here, not only physiologically moving organs are present, but also a lot of soft-tissue non-rigid organs which move in a very complex manner are present. Novel motion compensation algorithms must be integrated into the reconstruction, mostly non-rigid registration and fusion algorithms [35] to overcome these challenges.

**Assessing the Risk/Benefit Ratio of 4D Intervention Guidance**

As with all novel tools, the ratio between positive and negative effects must be carefully evaluated. This also holds true for 4D intervention guidance.

The advantages are clear. 4D intervention guidance may make interventions safer, less invasive, and may reduce the cost of an intervention. Furthermore, interventions may become quicker, while more complex interventions may also become possible.

Ultimately, it might be deemed wishful thinking to want to analyze the risk-benefit ratios by comparing large, prospective clinical studies with considerable follow-up evaluations. However, until such studies are conducted, one must employ common sense, reasoning, and surrogate markers in order to evaluate the risk benefit ratio of 4D intervention guidance.

Possible standards for comparison are standard fluoroscopy or step-and-shoot CT intervention guidance, optimally outfitted with assistance systems such as navigation systems [8]. Dose evaluations should carefully select the appropriate relationship: A method that applies a higher dose on a per time basis might still be preferable because the overall intervention length is greatly reduced due to better guidance. Furthermore, the complication rate and the benefit of totally novel interventions might well justify an increase in radiation dose.

Another problem of finding useful standard comparison are different dose measurement methods that are in use with fluoroscopy and tomography [36]. To compare effective dose on a per procedure base would be most preferable. However, in order to calculate the effective dosage, one must know the local dose distribution within the patient and organs. Considerable difficulties are tied to this in reality. One example is when substitute measurements are recorded, which are different in fluoroscopic and tomographic imaging, aggravating the direct comparison of both intervention guidance methods: The dose area product is the method of choice in fluoroscopy, while CTDI is the standard in tomographic imaging. Both measurements cannot easily be converted without in-depth knowledge of the precise imaging geometry and radiated anatomy. An alternative, feasible method to generate a measurement which is good for a direct comparison would be to record the dose area product in tomographic scanning and then sum it up with the fluoroscopy. Another way would be to scan a phantom using both methods and use it as an internal standard for relative assessments [2].

**A New challenge for Machine-Human Interface: Realtime Display of 4D Information for Intervention Guidance**

To generate a continuous, real-time data stream that depicts intervention instruments and anatomy in a spatial manner is only one side of the coin. This data stream must be displayed in a manner from which the interventionalist can derive useful information. Standard post-processing techniques are available to display the 3D information, including multiplanar reformation, volume-rendering, segmentation, and so on. An interesting feature is the calculation of virtual X-ray sum images that mimic projection fluoroscopy images. If generated in an acceptable time resolution with sufficient image quality, these images can provide impressions comparable to standard fluoroscopy.

The novel challenges to display the current data stream in intervention guidance are doing so with minimal user interaction and in real-time. If one pictures the time that is needed to manually generate a meaningful 3D display from a 3D dataset in a purely diagnostic setting, including finding a good volume rendering transfer function, the right selection of the volume of interest, or segmentation, one can imagine that doing so in real-time and without any user interaction is a significant challenge. Since this necessity has never before occurred, some significant research is needed. To address this, standard tools such as volume rendering, multiplanar reformations, segmentations, and so on can be combined with automatic region of interest detection (e.g. most prominent changes in the volume representing movement of instruments) and high level knowledge or model based approaches. User interaction is possible, but it must be reduced to a minimal level in order to leave the interventionalist free to actually perform the intervention. It would be acceptable to manually define a region of interest once (e.g. in while coiling an aneurysm) and to select the best view angles while algorithms do the rest.

**Discussion: Gantry Based Cath Lab Interventions ?**

4D intervention guidance requires the acquisition of repeatedly acquired, tomographic datasets. Hence, in the intervention suite it needs at least some sort of evolving system. It is not necessarily thought that this needs to be via full 360° rotations or large arc movements. A small-arc, forward/backward 'wobbling' system might be sufficient to sample the necessary data. Open systems such as C-arm based systems are prone to interruption if hardware enters its movement trajectory. The rotation speed is limited below that technically possible so as to protect personnel and patients. This is currently not a problem because tomographic datasets are, at most, acquired only a couple of times during an intervention, and personnel can take necessary precautions (they might leave the intervention suite anyway due to radiation protection). However, with respect to future developments in 4D intervention guidance, the rotation speed becomes a limiting factor to the time resolution. A fast evolution is necessary to scan enough data for a given time resolution. Time resolution should be considerably subsecond and hence full evolution times must be subsecond. Multi-threaded (e.g. biplane) systems can reduce the evolution speed, but still, if a high temporal resolution is needed, there will be a quickly revolving or `wobbling` system in the intervention room. Taking all this into

---

account C-arm setups are therefore limited with respect to the time resolution, making an argument for a gantry-based approach.

Another argument for gantry-based setups is the following: During more complex interventions, the intervention room is usually full of cables, oxygen pipes, lines, and instruments, not to mention the input consoles, monitors, and other measurement systems. They are covered with aseptic covers that must not have contact with septic surfaces (Fig. 4). To acquire a CT scan, the C-arm must be brought into a certain position which is usually much different from the position which would have previously been used to guide the intervention. Many instruments, lines, and pipes block the pathway of the C-arm. Much time must be spent in rearranging the setup and clearing the path. All of this increases the time and effort needed to acquire a CT scan. Therefore, this increases the threshold for employing the CT as a useful tool during interventions.

A gantry-based setup overcomes these limitations. The rotation pathway can never be blocked because the housing will protect it. Obviously, instruments, lines, and cables do not have to be rearranged in order to acquire a CT scan. A CT scan could be acquired almost immediately at a push of a button, making the 3D information much more accessible.

A rotation speed of up to almost 3 rotations per second is technically possible and it would provide a temporal resolution of up to 6 novel and independent datasets per second (Roughly a 180° rotation is need to reconstruct a novel, tomographic data set).

Wide-bore intraprocedural CT scanners have recently been introduced into the market [37]. However, they are currently unable to rotate continuously and/or transfer data through a live link for real-time reconstruction.

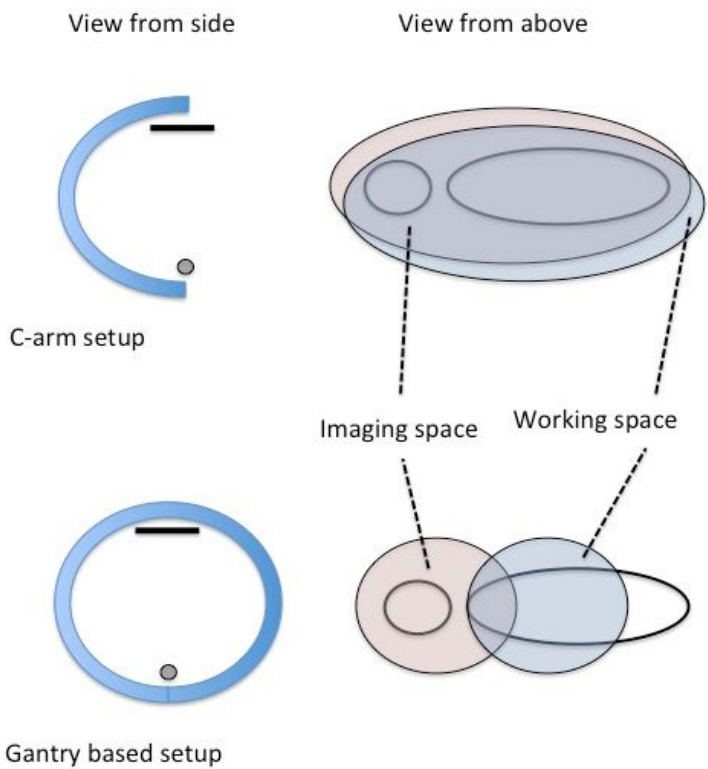

**Figure 7: Gantry based intervention guidance vs. C-arm based setups**: C-arm based set ups are considered to be the current state-of-the-art for angiographic interventions. However, gantry-based approaches might become justified as the value of (continuous) 3D information becomes more and more immanent, even in the realm of cath lab interventions where the difference in patient access might require relevant workflow changes.

Concerns with respect to gantry-based systems are:

1. Limited access to the patient (Fig. 7) - the

    interventionalist cannot work next to the gantry.
2. Reduced degrees of freedom in finding appropriate fluoroscopy projection positions - a gantry would allow to tilt the imaging chain in one plane (the rotating plane), tilting in the other planes would be very limited if possible at all.
3. Lack of the ability to perform certain interventions - fluoroscopically guided PTCD or drainages would be difficult, because they would require the interventionalist to stand in the imaging plane.

While these concerns are legitimate, they can be refuted because of their strengths in the following counter points:

Ad 1: Novel, gantry based cone-beam CT setups designed for intraoperative use, demonstrate that today's gantries can be built with a very slim design and a rather larger bore. Thereby they are limiting access to the patient considerably less compared to older, diagnostic CT gantries. Most interventions do not need access to the patient on the direct imaging plane. E.g. most vascular interventions are conducted from the groin and the relevant volume of interest is at a location that is far from the access site. In situations in which imaging next to the introduction sheath would be necessary, e.g. in case of complications, a sufficiently large bore will grant enough access. Furthermore, alternative accesses to the vasculature such as through the brachial vessels can be used.

Ad 2: A gantry based setup will certainly limit the ability to freely approach projection angulations in contrast to C-arm setups. However, even in a gantry based setup, one degree of angulation remains accessible without limitations; within the gantry plane, every projection position can be reached and selected. Angulating off of the gantry plane is certainly limited, but still feasible, e.g. the gantry as a whole could be tilted. However, this will certainly be less practicable than angulating using a freely moving C-arm setup. One question remaining is whether it will be relevant to freely select projection positions in the future? The ability to acquire CT scans easily and quickly might very well overcome certain limitations potentially occurring from gantry based setups. Furthermore, it offers the ability to reconstruct 3D datasets in swift succession or the ability to reconstruct virtual projections from every angle with 4D datasets.

Ad 3: A gantry based setup might issue procedural changes in how interventions may be conducted, but most likely these changes will not result in the loss of the ability to conduct certain procedures. Rather, the way these procedures will be conducted will change; e.g. already at present, percutaneous drainages of the gall system can be carried out under fluoroscopic guidance, as well as CT-guidance, alone. One may easily envision as to how a combination of both techniques, fluoroscopy as well as CT, can fully satisfy all guidance needs, even if the freedom to find appropriate positions for the fluoroscopy is somewhat limited.

# Translational Outlook and Clinical Introduction

The increasing use of cone-beam CT for image guidance is very much a reality in daily clinical practice, however, 4D intervention guidance is not. The developmental status is 'proof-of-concept in phantom and animal studies'. Many unaddressed problems exist. However, none of these seem to be indicating that currently unforeseeable discoveries need to be assumed in order to overcome them. Incorporating more and more complex postprocessing algorithms into tomographic reconstruction as well as external knowledge, etc. might well afford greater and greater savings in radiation dose. Other fields in which lots of graphical as well as structural data are being processed and brought into a meaningful context, e.g. computer games, autonomous driving, etc. seem to be much more evolved. Intervention guidance might well benefit from their attainments once the industry can justify knowledge transfer in between these very different markets.

Even if full fledged 4D intervention guidance will not be reached in the short or mid term in clinical practice, algorithms and methods which are being developed in this context might be used even in current setups in order to realize dose reductions in repetitive CT scans that might be taken at different points in time during the intervention.

Furthermore, the herein discussed reconstruction algorithms can also be utilized in other, much different scanning hardware setups. Multi-threaded systems with many more than a few imaging chains might provide other interesting characteristics and trade-offs. This holds especially true, if X-ray sources and detectors are at least partially stationary and/or independently rotating [38].

**First Use Case for 4D Intervention Guidance: Neurointerventions**

Catheter procedures in the head and neck region seem to be best suited for 4D intervention guidance. Many neurointerventions are highly complex, and the need for the depiction of an accurate 3D relationship of intervention instruments (e.g. stent struts / coils) as well as the underlying anatomy (e.g. aneurysm sacks) is highly appreciated [23]. Furthermore, on the technical side, the head region, with its relative small scan field-of-view as well as the lack of moving organs, seems to be the least demanding area to implement 4D intervention guidance: Its scan field-of-view geometry fits standard flat-panel detectors. A gantry-based setup seems to need the least amount of reorganization in workflows because the imaging space (head) is in the large majority of neuro-interventions far away from the working space (groin). Specialized neuro-interventional C-arm setups are being marketed to the neurointerventional community, so a specialized market already exists. This market segment could be the first for 4D intervention guidance.

## In Conclusion

This overview was stipulated by the prospects of truly 4D intervention guidance using X-rays and its huge implications for minimal-invasive medicine. The real accomplishment of this work should be to point the research community in the direction of X-ray based intervention guidance to achieve full 4D intervention guidance. The research community should foster research in this field with an objective to combine novel reconstruction algorithms together with smart post-processing and new data acquisition strategies. It should consider a break from the current mantra of `C-arm equals intervention guidance`, and so, it should consider gantry-based catheter intervention approaches as a full alternative to C-arm setups.

In past years, much effort was spent on other guidance methods which are not being introduced into clinical medicine at all at a relevant scale. One example is MRI guidance that does not play a relevant role in clinical medicine even after decades of research. If only a small percentage of these research efforts were to be put into 4D intervention guidance using X-rays, the chances are that minimal-invasive medicine would have an affordable 4D guidance method available within years from now.

## Acknowledgment

Sedat Aktas, X-ray imaging and computed tomography, Medical physics in Radiology, German Cancer Research Center, for image reconstruction and processing. Dr. Nils Rathmann, Department of Diagnostic imaging and Nuclear Medicine, Mannheim University Medical Center, for intraprocedural intervention pictures.

# References

1.  Kuntz J, Gupta R, Schönberg SO, Semmler W, Kachelrieß M, Bartling S. Real-time X-ray-based 4D image guidance of minimally invasive interventions. Eur Radiol. 2013;23: 1669–1677. doi:10.1007/s00330-012-2761-2

2.  Kuntz J, Flach B, Kueres R, Semmler W, Kachelrieß M, Bartling S. Constrained reconstructions for 4D intervention guidance. Phys Med Biol. 2013;58: 3283–3300. doi:10.1088/0031-9155/58/10/3283

3.  Flach B, Kuntz J, Brehm M, Kueres R, Bartling S, Kachelrieß M. Low dose tomographic fluoroscopy: 4D intervention guidance with running prior. Med Phys. 2013;40: 101909. doi:10.1118/1.4819826

4.  Kos S, Huegli R, Bongartz GM, Jacob AL, Bilecen D. MR-guided endovascular interventions: a comprehensive review on techniques and applications. Eur Radiol. 2007;18: 645–657. doi:10.1007/s00330-007-0818-4

5.  Bock M, Wacker FK. MR-guided intravascular interventions: techniques and applications. J Magn Reson Imaging. 2008;27: 326–338. Available: http://www.ncbi.nlm.nih.gov/entrez/query.fcgi?cmd=Retrieve&db=PubMed&dopt=Citation&list_uids=18219686

6.  Salamon J, Hofmann M, Jung C, Kaul MG, Werner F, Them K, et al. Magnetic Particle / Magnetic Resonance Imaging: In-Vitro MPI-Guided Real Time Catheter Tracking and 4D Angioplasty Using a Road Map and Blood Pool Tracer Approach. PLoS One. 2016;11: e0156899. doi:10.1371/journal.pone.0156899

7.  Wood BJ, Zhang H, Durrani A, Glossop N, Ranjan S, Lindisch D, et al. Navigation with Electromagnetic Tracking for Interventional Radiology Procedures: A Feasibility Study. J Vasc Interv Radiol. 2005;16: 493–505. doi:10.1097/01.RVI.0000148827.62296.B4

8.  Kickuth R, Reichling C, Bley T, Hahn D, Ritter C. C-Arm Cone-Beam CT Combined with a New Electromagnetic Navigation System for Guidance of Percutaneous Needle Biopsies: Initial Clinical Experience. Rofo. 2015;187: 569–576. doi:10.1055/s-0034-1399313

9.  Kim S-H, Oh Y-S, Kim D-H, Choi IJ, Kim T-S, Shin W-S, et al. Long-term outcomes of remote magnetic navigation for ablation of supraventricular tachycardias. J Interv Card Electrophysiol. 2015;43: 187–192. doi:10.1007/s10840-015-9991-6

10. Stangenberg L, Shuja F, Carelsen B, Elenbaas T, Wyers MC, Schermerhorn ML. A novel tool for three-dimensional roadmapping reduces radiation exposure and contrast agent dose in complex endovascular interventions. J Vasc Surg. doi:10.1016/j.jvs.2015.03.041

11. Sailer AM, Haan MW de, Graaf R de, Zwam WH van, Schurink GWH, Nelemans PJ, et al. Fusion Guidance in Endovascular Peripheral Artery Interventions: A Feasibility Study. Cardiovasc Intervent Radiol. 2014;38: 314–321. doi:10.1007/s00270-014-0951-9

12. Byrne JV, Colominas C, Hipwell J, Cox T, Noble JA, Penney GP, et al. Assessment of a

technique for 2D–3D registration of cerebral intra-arterial angiography. BJR Suppl. Februar 1, 2004;77: 123–128. doi:10.1259/bjr/27339681

13. Volpi D, Sarhan MH, Ghotbi R, Navab N, Mateus D, Demirci S. Online tracking of interventional devices for endovascular aortic repair. Int J CARS. 2015;10: 773–781. doi:10.1007/s11548-015-1217-y

14. E K Paulson DHS. CT fluoroscopy--guided interventional procedures: techniques and radiation dose to radiologists. Radiology. 2001;220: 161–167. doi:10.1148/radiology.220.1.r01jl29161

15. Buls N, Pagés J, de Mey J, Osteaux M. Evaluation of patient and staff doses during various CT fluoroscopy guided interventions. Health Phys. 2003;85: 165–173. Available: http://www.ncbi.nlm.nih.gov/pubmed/12938963

16. ICRP, Rehani MM, Gupta R, Bartling S, Sharp GC, Pauwels R, et al. Radiological Protection in Cone Beam Computed Tomography (CBCT). ICRP Publication 129. Ann ICRP. 2015;44: 9–127. doi:10.1177/0146645315575485

17. Schegerer AA, Lechel U, Ritter M, Weisser G, Fink C, Brix G. Dose and image quality of cone-beam computed tomography as compared with conventional multislice computed tomography in abdominal imaging. Invest Radiol. 2014;49: 675–684. doi:10.1097/RLI.0000000000000069

18. Wallace MJ, Kuo MD, Glaiberman C, Binkert CA, Orth RC, Soulez G. Three-dimensional C-arm cone-beam CT: applications in the interventional suite. J Vasc Interv Radiol. 2008;19: 799–813. Available: http://www.ncbi.nlm.nih.gov/entrez/query.fcgi?cmd=Retrieve&db=PubMed&dopt=Citation&list_uids=18503893

19. Tacher V, Radaelli A, Lin M, Geschwind J-F. How I Do It: Cone-Beam CT during Transarterial Chemoembolization for Liver Cancer. Radiology. Januar 27, 2015;274: 320–334. doi:10.1148/radiol.14131925

20. Schwartz JG, Neubauer AM, Fagan TE, Noordhoek NJ, Grass M, Carroll JD. Potential role of three-dimensional rotational angiography and C-arm CT for valvular repair and implantation. Int J Cardiovasc Imaging. 2011;27: 1205–1222. doi:10.1007/s10554-011-9839-9

21. Thakor AS, Chung J, Patel R, Cormack R, Legiehn G, Klass D. The use of cone-beam CT in assisting percutaneous translumbar catheter placement into the inferior vena cava. Clin Radiol. 2015;70: 21–24. doi:10.1016/j.crad.2014.09.009

22. Kamran M, Nagaraja S, Byrne JV. C-arm flat detector computed tomography: the technique and its applications in interventional neuro-radiology. Neuroradiology. 2009;52: 319–327. doi:10.1007/s00234-009-0609-5

23. Kuriyama T, Sakai N, Niida N, Sueoka M, Beppu M, Dahmani C, et al. Dose reduction in cone-beam CT scanning for intracranial stent deployment before coil embolization of intracranial wide-neck aneurysms. Interv Neuroradiol. 2016;

doi:10.1177/1591019916632489

24. Pjontek R, Önenköprülü B, Scholz B, Kyriakou Y, Schubert GA, Nikoubashman O, et al. Metal artifact reduction for flat panel detector intravenous CT angiography in patients with intracranial metallic implants after endovascular and surgical treatment. J Neurointerv Surg. 2015; doi:10.1136/neurintsurg-2015-011787

25. Pan X, Sidky EY, Vannier M. Why do commercial CT scanners still employ traditional, filtered back-projection for image reconstruction? Inverse Probl. 2009;25: 1230009. doi:10.1088/0266-5611/25/12/123009

26. Marcel Beister DK. Iterative reconstruction methods in X-ray CT. Phys Med. 2012;28: 94–108. doi:10.1016/j.ejmp.2012.01.003

27. Willemink MJ, Jong PA de, Leiner T, Heer LM de, Nievelstein RAJ, Budde RPJ, et al. Iterative reconstruction techniques for computed tomography Part 1: Technical principles. Eur Radiol. 2013;23: 1623–1631. doi:10.1007/s00330-012-2765-y

28. Dang H, Stayman JW, Sisniega A, Xu J, Zbijewski W, Wang X, et al. Statistical reconstruction for cone-beam CT with a post-artifact-correction noise model: application to high-quality head imaging. Phys Med Biol. 2015;60: 6153–6175. doi:10.1088/0031-9155/60/16/6153

29. Tilley S, Siewerdsen JH, Stayman JW. Model-based iterative reconstruction for flat-panel cone-beam CT with focal spot blur, detector blur, and correlated noise. Phys Med Biol. 2016;61: 296–319. doi:10.1088/0031-9155/61/1/296

30. Kaganovsky Y, Li D, Holmgren A, Jeon H, MacCabe KP, Politte DG, et al. Compressed sampling strategies for tomography. J Opt Soc Am A. Juli 1, 2014;31: 1369–1394. doi:10.1364/JOSAA.31.001369

31. Jørgensen JS, Sidky EY. How little data is enough? Phase-diagram analysis of sparsity-regularized X-ray CT. arXiv:14126833 [cs, math]. 2014; Available: http://arxiv.org/abs/1412.6833

32. Sidky EY, Chartrand R, Boone JM, Xiaochuan Pan. Constrained ${\rm T}p{\rm V}$ Minimization for Enhanced Exploitation of Gradient Sparsity: Application to CT Image Reconstruction. IEEE J Transl Eng Health Med. 2: 1–18. doi:10.1109/JTEHM.2014.2300862

33. Park C, Zhang H, Chen Y, Fan Q, Kahler D, Li J, et al. WE-G-207-03: Mask Guided Image Reconstruction (MGIR): A Novel Method for Ultra-Low-Dose 3D and Enhanced 4D Cone-Beam Computer-Tomography. Med Phys. American Association of Physicists in Medicine; 2015;42: 3696–3696. doi:10.1118/1.4926096

34. Qi Z. TH-CD-303-08: A Novel 4D CBCT Reconstruction Method Using Registraition Assisted Compressed Sensing (RACS). Med Phys. American Association of Physicists in Medicine; 2015;42: 3730–3730. doi:10.1118/1.4926243

35. Flach B, Brehm M, Sawall S, Kachelrieß M. Deformable 3D-2D registration for CT and its

application to low dose tomographic fluoroscopy. Phys Med Biol. 2014;59: 7865–7887. doi:10.1088/0031-9155/59/24/7865

36. Abuhaimed A, Martin CJ, Sankaralingam M, Gentle DJ. Investigation of practical approaches to evaluating cumulative dose for cone beam computed tomography (CBCT) from standard CT dosimetry measurements: a Monte Carlo study. Phys Med Biol. 2015;60: 5413. doi:10.1088/0031-9155/60/14/5413

37. Hecht N, Kamphuis M, Czabanka M, Hamm B, König S, Woitzik J, et al. Accuracy and workflow of navigated spinal instrumentation with the mobile AIRO(®) CT scanner. Eur Spine J. 2015; doi:10.1007/s00586-015-3814-4

38. Besson GM. New CT system architectures for high temporal resolution with applications to improved geometric dose efficiency and cardiac imaging. Med Phys. 2015;42: 2668–2678. doi:10.1118/1.4918328
Bartling S, Kachelrieß M, 2016, *X-ray tomographic intervention guidance: Towards real-time 4D imaging,* Preprint

22